\documentclass[12pt,a4paper]{article}
\usepackage[T2A]{fontenc}
\usepackage[cp1251]{inputenc}
\usepackage[english]{babel}
\usepackage{amssymb}
\usepackage{amsmath}
\usepackage{graphicx}
\usepackage[small,sc,center]{titlesec}
\usepackage{indentfirst}
\usepackage[labelsep=period]{caption}
\usepackage{caption}
\newcommand{\msun}{\,$M_{\odot}$}
\newcommand{\msyr}{\,$M_{\odot}$\,yr$^{-1}$}
\newcommand{\rsun}{\,$R_{\odot}$}
\newcommand{\ergs}{\,erg\,s$^{-1}$}

\newcommand{\gcm}{\,g\,cm$^{-1}$}
\newcommand{\kms}{\,km\,s$^{-1}$}

\newcommand{\ha}{H$\alpha$}

\begin{document}

\begin{center}
	
\textbf{\large Supernova 2009ip outbursts in 2012: From scenario to model} 

\vskip 5mm
\copyright\quad
2022 г. \quad N. N. Chugai\footnote{email: nchugai@inasan.ru} \\
\textit{$^1$Institute of astronomy, Russian academy of science, Moscow} \\

\end{center}


\noindent
{\em PACS codes:\/} 

\vskip 2cm

 \begin{abstract} 
 	
Spherical and aspherical models are presented for two outbursts in 2012 
of supernova 2009ip. Models are based on a scenario which suggests that the 
August 2012 outburst is caused by the explosive shell ejection from LBV-presupernova. 
The model predicts an emergence of an unobserved outburst in late July 2012 related to 
a shock breakout and a subsequent diffusive radiative cooling of the ejected envelope. 
The luminosity of the first observed outburst in August 2012 was presumably powered by the central 
source, whereas the second, more powerful outburst in late September 2012, was caused by the ejecta interaction with the circumstellar envelope.
Models provide estimates of the ejecta energy and mass along with the mass of the circumstellar shell.

\end{abstract}

\clearpage

\section{Introduction}

Supernova SN~2009ip in NGC7259 galaxy is a complex phenomenon following  
the 20/08/2009 outburst with the absolute magnitude of -14.5 
(Maza et al. 2009). Presupernova is found in archive image of NGC7259 taken by 
the Hubble telescope ({\em HST} hereafter) with the absolute magnitude of -9.5 and 
was identified as LBV (Smith et al. 2010). Two years after the discovery SN~2009ip experienced several 
outbursts in the range of -10.5... -13.5 in the span of 160 days (Pastorello et al. 2013).

Key events happened after 2012 August 8, when SN~2009ip experienced outburst 
up to -15 mag (aka 2012a outburst) followed by the late September, 2012 powerful 
outburst (2012b outburst) up to -18 mag (Pastorello et al. 2013) with the subsequent 
monotonic flux decrease. In December 2021 the absolute magnitude in F606W filter 
of {\em HST} is found to be -8.5, one magnitude fainter compared to -9.5 mag in 1999, so authors concluded that the LBV-presupernova has gone (Smith et al. 2022).

Starting with the discovery spectra have been showing strong narrow hydrogen emission lines with the FWHM of 800-1000\kms\ (Pastorello et al. 2013, Margutti et al. 2014) indicating heavy mass loss. The first spectrum 09/08/2012 of the 2012a outburst 
revealed broad hydrogen P Cyg line profiles with wing velocities up to 14000\kms\ (Pastorello et al. 2013). This observation argued a scenario in which the 
2012a outburst was the supernova explosion, whereas the subsequent 
more powerful outburst 2012b was caused by the interaction of the supernova ejecta with the circumstellar envelope (Mauerhan et al. 2013, Pastorello et al. 2013).
An alternative possibility suggests two explosive events related respectively with 2012a and 2012b outbursts with the subsequent CS interaction after the second explosion that presumably occured 13/09/2012 (Margutti et al. 2014). Details of this 
scenario are bounded by analytical estimates. For the first, single explosion scenario, the light curve modelling and estimates of supernova energy and ejected 
mass are absent at all. 
Yet this scenario deserves detailed study since it requires only one explosion.

The identification of the SN~2009ip presupernova with the massive LBV implies that the observed phenomena in this case are related to nuclear burning flashes in oxygen core initiated by the instability cased by pair creation (Woosley 2017, 
Smith et al. 2022). Supernovae related to such phenomena are called 
pulsational pair-instability supernovae (PPISN for short).
Yet one has to emphasise that full resemblance between SN~2009ip 
and PPISN models (Woosley 2017) is lacking. The crucial mismatch is the point that 
SN~2009ip spectra during 2012a outburst show high expansion velocity up to 
13000-14000\kms, which is 5-10 larger compared to velocities of ejected shell in
PPISN models.  

The paper addresses the scenario of a single explosion of a LBV related to the 2012a outburst of SN~2009ip.
The presented model suggests explosive shell ejection (2012a outburst) and its subsequent 
interaction with the CS envelope (2012b outburst). Given significant polarization of 
SN 2009ip radiation (Mauerhan et al. 2014) we explore models with spherical and aspherical distribution of the CS matter.

\section{Light curve modelling}

\subsection{Preliminary consideration}

Based on the premis that 2012a outburst is caused by the explosive mass ejection by the 
LBV one should adopt the presupernova radius of 70\rsun\ in line with the evolutionary models of LBV with masses of 80-90\msun (Woosley 2017). With such a radius the initial 
luminosity related to the shock breakout (call it "2012a0" outburst) can exceed the luminosity of 2012a outburst. Moreover, since the initial luminosity peak of 2012a0 outburst 
and the subsequent light curve has to differ from the light curve of 2012a, one should 
admit that the initial 2012a0 outburst was missing, so the supernova explosion occured somewhat earlier than turn on of the 2012a outburst. 
Yet the explosion should be later than the flash of 24/07/2012 at the level of 
$V \approx 17.4$ (Drake et al. 2012). 

Along with the explosive ejection of the supernova (2012a outburst), as we see below,
 one has to admit additional thermal energy injection into the expanding envelope 
 of the order of $\sim 2\times10^{48}$ erg. The supernova light curve will be calculated 
 based on the analytical model of the radiation diffusion with the adiabatic cooling and 
the additional source of the thermal energy (Arnett 1982). The subsequent CS interaction 
is treated based on the thin shell approximation (Giuliani 1982, Chevalier 1982).
The detailed realization of the CS interaction light curve model is described earlier 
(Chugai 2001, 2021). We therefore omit model description, but remind that the supernova 
shell deceleration in the dense CS gas proceeds via formation of outer (direct) and reverse shock waves with the formation of a cold dense shell (CDS) inbetween.
The CDS can be optically thick and in this case the supernova shows smooth continuum without broad absorption lines; this phenomenon took place in SN~1998S first month after explosion (Chugai 2001).

We consider spherical and aspherical options for the CS mater distribution. 
Non-sphericity is supported by the significant polarization of the SN~2009ip radiation 
during 2012b outburst (Mauerhan et al. 2014).
Non-spherical model will be calculated based on the spherical model with the 
CS interaction luminosity reduced by an asphericity factor being equal to the ratio of 
a visible area of the aspherical-to-spherical photosphere: $\xi = A_{ns}/A_s < 1$.
To a first approximation $\xi$ is equal to the ratio of radii squared of the 
photometric photosphere (Margutti et al. 2014) and the photosphere of a spherical model. 
In fact, the photometric radius determined based on the blackbody approximation is lower than the geometric value
 because the escaping flux from the atmosphere with the absorption 
and scattering is lower than blackbody flux by a factor of 
 $\sqrt{\epsilon}/(1+\sqrt{\epsilon}) < 1$, where $\epsilon$ is the photon loss probability 
 per one act of extinction (cf. Mihalas 1978).
 It should be emphasised that in aspherical model the supernova envelope is 
 assumed to be spherical; asphericity refers to the CS shell only.
 
Preliminary modelling of the 2012b outburst indicates that a steep luminosity 
rise requires the steep density drop in outer layers ofthe  supernova shell $\rho \propto 
v^{-\omega}$ with $\omega \geq 20$. The density in the inner layers are assumed to be 
uniform. The CS shell should have a sharp inner boundary as well to provide the steep luminosity 
rise. The inner zone of the CS medium is presumably produced by a steady wind 
of a moderate density with the density parameter $w = \dot{M}/u = 10^{14}$\gcm\ that corresponds to the mass 
loss rate of $1.5\times10^{-4}$\msyr\ for the speed of 1000\kms. The dense CS shell presumably flows with the 
same speed of $u = 1000$\kms. 
For the CS shell mass of $\sim 1$\msun\ its Thomson optical depth 
in the model is less than unity. which permits us to neglect by the radiation diffusion delay compared to the light 
travel time $r/c$ and adopt the approximation of the 
instant radiation escape. The modelling shows that the light travel effect does not modify 
significantly the light curve of 2012b outburst as well.

Model parameters (kinetic energy and mass of supernova ejecta, radius and mass of the CS envelope) are constrained by the 
light curve and maximal velocity of the undisturbed supernova ejecta that are determined from a blue wing 
of the absorption component of  \ha, H$\beta$ at the 2012a-stage  and He\,I 10830\,\AA\ at 2012b-stage.

\begin{table}[t]
	\vspace{6mm}
	\centering
	{{\bf Table 1} Parameters of supernova and CS envelope}
	
	\vspace{5mm}\begin{tabular}{c|c|c|c|c|c|c|c} 
		\hline	
		&     &   &  & &   &  & \\
  Model  & $E_{51}$~$^a$  & $M/M_{\odot}$ & $t_{inj}$ (d) & $L_{0,40}$~$^b$ & $M_{cs}/M_{\odot} $ & $r_{1,15}$~$^c$ & $r_{2,15}$ \\
		\hline
  sm  &  0.34   &  0.2  & 24  & 90   & 1.3      & 8.4  & 13 \\  
  nsm  &  2     &  0.8  & 22   & 110 & 3.2  &  8.8 &  13.4 \\  	

		\hline
		
	\end{tabular}
\parbox[]{10cm}{ \small {Units: $^a$ $10^{51}$\,erg; $^b$ $10^{40}$\ergs;  
		$10^{15}$\,cm }}	
\end{table}

\subsection{Modelling results}

The supernova light curve description (2012a outburst), as noted earlier, requires  a slow post-explosion 
injection of the thermal energy. The temporal behavior of the power 
released by a central source is adopted as 
\begin{equation}
L = L_0(t/t_{inj})^p\exp{[-(t/t_{inj})^2)]}\,,
\end{equation} 
where the power index $p \sim 1.5-2$, while $t$ is the time lapsed from the adopted explosion moment JD 2456133.5 (25/07/2012).  

First month after the light maximum of the 2012b outburst the CDS is optically thick as 
indicated by the smooth continuum. Since the ejecta mass and the CDS optical depth 
decreases as the kinetic energy decreases, one expects the minimal acceptable energy. 
The modelling indicates that the kinetic energy cannot be lower than $ 4\times10^{50}$\,erg. For the spherical fiducial
 model (model sm, Table 1, Figure 1) we adopt $E = 3.4\times10^{50}$\,erg.
For the aspherical model (nsm, Table 1, Figure 2) with the asphericity parameter 
$\xi = 0.05$ the minimal energy is significantly higher, $E = 2\times10^{51}$\,erg.
The Table 1 includes supernova ejecta energy and mass, the timescale and power of 
the additional energy injection, CS mass, the inner and outer radius of the CS shell.  

Both models describe the bolometric light curves of 2012a and 2012b outbursts and 
are consistent with the observational maximal velocities of unshocked supernova ejecta 
at about 100 days when CDS becomes transparent in the continuum.
Noteworthy, the model expansion velocity of the undisturbed supernova envelope can 
be larger, but cannot be lower, than the maximum velocity recovered from broad absorption lines. 

At the very late phase (400 days) in the Keck-2 spectrum (Graham et al. 2017) the velocity estimated from 
wings of emission lines of  H$\beta$, He\,I 7065\,\AA, and Fe\,II 5018\,\AA, 
are equal to 1500-2000\kms\ that is lower than the CDS velocity and maximal velocity of the undisturbed
 supernova shell in the spherical model (Figure 1). In a context of the spherical model this suggests 
 that the contribution of the CDS and unshocked ejecta in these lines 
is relatively small, whereas the dominant source of emission lines might be shocked dense CS clouds following a scenario for the intermadiate component of emission lines 
in SN~1988Z (Chugai \& Danziger 1994). In the aspherical model, however, the CDS velocity 
on day 400 ($\approx 1400$\kms, Figure 2) is close to the emission lines FWHM, so the CDS could be the emitting site for these lines.

An easy way to realise an aspherical model would be the supernova interaction with a CS 
shell arranged as a disk (torus). This structure is invoked, e.g., to account for 
the co-existence of broad emission lines and the intermediate component in SN~1988Z (Chugai \& Danziger 1994). 
A similar structure is considered to account for the polarization of the continuum 
originated presumably from the shocked interaction at the inner edge toroidal CS shell (Mauerhan et al. 2014).
In the aspherical model CS shell asphericity parameter $\xi = 0.05$ corresponds to 
the equatorial belt in the latitude range $\pm 3^{\circ}$ for the inclination angle of $90^{\circ}$.
The value $\xi = 0.05$ approximately corresponds to the ratio of radii squared of the 
photometric photosphere (Margutti et al. 2014) and the photosphere of spherical model with a possible correction on the lower surface brightness compared to the blackbody intensity. 

A high density contrast of the toroidal CS shell with respect to the supernova envelope 
result in a situation when almost unaccelerated CS torus submerges into the expanding supernova envelope. 
 Eventually we would obtain a picture in which the interaction region 
at the inner torus edge becomes an internal source for the continuum radiation on which background 
 broad absorption lines could form. On the other hand, the 
presence of absorption lines in the spectrum at the 2012b-stage is a serious problem for the spherical model which makes the aspherical model more advantageous.
 
Noteworthy, the postmaximum fast luminosity decline of 2012b outburst can be well 
described by the power law $L \propto t^{-6}$ that follows from a model of 
the shock interaction of homologously expanding envelope with the spherical shell 
of the radius $R$ at rest. Indeed, the incoming speed at the radius $R$ decreases 
with time as  $v \propto R/t$, whereas the density drops as $\rho \propto t^{-3}$, 
so the luminosity of the reverse shock is $L \propto R^2\rho v^3 \propto t^{-6}$. 
This simple model can reflect the supernova interaction with the dense toroidal CS shell that 
is weeakly accelerated by the rarefied flow of supernova ejecta.

\section{Discussion and conclusion}

The paper has been aimed at the modelling of two outbursts of SN~2009ip in 2012 based on 
a scenario of explosive shell ejection by LBV presupernova. The scrnario suggests 
that the supernova explosion corresponds to the relatively weak outburst in August 
2012 (2012a outburst), whereas powerful outburst of late September 2012 (2012b outburst) 
is caused by the shock interaction of supernova shell with the CS shell (Pastorello et al. 2013, Mauerhan et al. 2013). According to the model the explosive ejection of the envelope by the LBV presupernova brings about strong initial flash 2012a0, preceded 2012a, with the luminosity of $\sim 10^{42}$\ergs\ and the timescale of one week that is missing in observations.

The light curve of both outbursts 2012a and 2012b and the expansion velocity of the 
supenova envelope are reproducible in both spherical model and models with toroidal (disk) CS shell. However, if for the spherical model the energy $\sim 3\times10^{50}$\,erg is sufficient, the aspherical model, on the other hand,  requires substantially higher value, of $\sim 2\times10^{51}$\,erg that may be in discomfort with PPISN models (Woosley 2017). Summarizing, phenomenologically 
the aspherical model is preferred, although it poses a serious problem for the explosion theory. 

Two features of the presented models should be emphasised.
First, the model supernova light curve (2012a outburst, Figures 1, 2) requires the prolong additional thermal energy injection. 
This energy is presumably supplied by the continued nuclear burning flash. 
Second, the explosive ejection of the shell related to the 2012a outburst is preceeded by 
the initial week-long outburst after the shock breakout (2012a0 outburst). The existence 
of this outburst around 25/07/2012 cannot be verified by available observations. 

This initial outburst is a generic feature of the LBV scenario with the presupernova radius of 70\rsun\ and is appropriate for the  verification. 
In this regards one should be emphasise that for the presupernova with 
the low radius, say 10\rsun, the initial luminosity peak 2012a0 turns out to be rather weak, at the level of initial luminosity of 2012a outburst. Since the full resemblance between 
SN~2009ip and PPISN models doesn't take place, the case of low presupernova radius cannot be rulled out. This remark is directly related to the test of the LBV scenario based on the observational presence/absence of the initial 2012a0 peak.

Following the termination 2012a outburst the LBV with the presupernova mass reduced by 
$\sim 4$\msun\ due to supernova and CS shells probably experienced collapse into a black hole, since the {\em HST} photometry in December 2021 suggests that the LBV-presupernova has gone 
(Smith et al. 2022).
Since the presupernova mass according to the {\em HST} photometry in 1999 was estimated as  
50-80\msun\ (Smith et al. 2010) the black hole mass should be in the range of 
50-80\msun.

Noteworthy, SN~2009ip is only partially unique, primarily due to a long set of observational data.
As to the phenomenon of 2012a and 2012b outburst, the similar photometric and spectral behavior was observed for SN~2010mc (Ofek et al., Smith et al. 2014).
Yet it is premature to conclude on the full resemblance, since SN~2010mc spectra 
at the stage corresponding to 2012a outburst are lacking, so one cannot say, whether 
the supernova explosion occurred long before the major outburst.

\vspace{0.8cm}

This study was partially supported by RFBR and DFG, project number 21-52-12032.

\vspace{1.5cm}

\centerline{\bf References}
\bigskip

\begin{flushleft}
	
Arnett W. D., Astrophys. J. {\bf 253}, 785 (1982)\\
\medskip
Chevalier R. A., Astrophys. J. {\bf 259}, 302 (1982)\\
\medskip
Chugai N. N., Mon. Not. R. Astron. Soc. {\bf 508}, 6023 (2021)\\
\medskip
Chugai N. N,, Mon. Not. R. Astron. Soc. {\bf 326}, 1448 (2001)\\
\medskip 
Chugai N. N., Danziger I. J.,  Mon. Not. R. Astron. Soc., 268, 173 (1994)\
\medskip
Drake A. J., ATel 4334 (2012)\\
\medskip
Graham M. L., Bigley A., Mauerhan J. C. et al.,  Mon. Not. R. Astron. Soc. {\bf 469}, 1559 (2014)\\                         
\medskip	
Giuliani J. L.,  Astrophys. J. {\bf 256}, 624 (1982) \\
\medskip
Margutti R., Milisavljevic D., Soderberg A. M. et al., Astrophys. J. {\bf 780}, 21 (2014)\\ 
\medskip
Mauerhan J., Williams G., Smith N. et al., Mon. Not. R. Astron. Soc. {\bf 442}, 1166 (2014)\\
\medskip
Mauerhan J., Smith N., Filippenko A. V. et al., Mon. Not. R. Astron. Soc. {\bf 430}, 1801 (2013)\\
\medskip
Maza J. et al., Cent. Bur. Electron. Telegrams, 1928, 1 (2009)\\
\medskip
Mihalas D. {\em Stellar armospheres} (San Francisco: Freeman and company, 1978)\\
\medskip 
Ofek E. O., Sullivan M., Cenko S. B. et al., Nature, {\bf 494}, 66 (2013) 
\medskip
Pastorello A., Cappellaro E., Inserra C., et al., Astrophys. J. {\bf 767}, 1 (2013)\\
\medskip
Smith N., Andrews J., Filippenko A. V. et al., Mon. Not. R. Astron. Soc. {\bf 515}, 715 (2022)\\
\medskip
Smith N., Mauerhan J. C., Prieto J. L., Mon. Not. R. Astron. Soc. {\bf 438}, 1191 (2014)\\
\medskip
Smith N., Miller A., Li W., Filippenko A. V. et al., Astron. J. {\bf 139}, 1452 (2010)\\
\medskip
Woosley S. E., Astrophys. J. {\bf 836}, 244 (2017)\\

\end{flushleft}
\clearpage

\clearpage

\begin{figure}
\centering
\includegraphics[width=0.97\columnwidth]{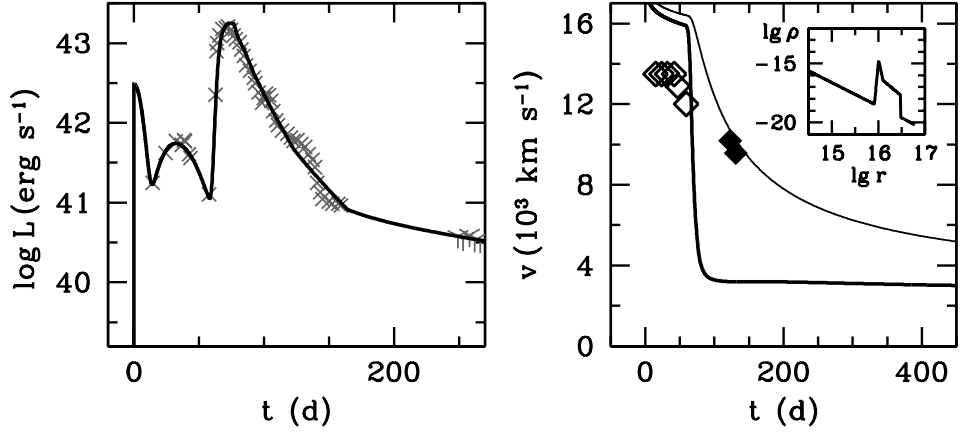}
\caption{\rm
{\em Left.} Bolometric light curve ({\em thick line}) in the spherical model (model sm, Table 1) compared to the observational light curve ({\em crosses} from Margutti et al. 2014; {\em Y-symbols} from Graham et al. 2017). First unobserved  maximum corresponds to the diffusion 
radiation cooling of supernova after the shock breakout, the second is the 2012a outburst 
that reflects the supernova emission powered by the additional energy release by the central source, the third maximum is the result of the supernova interaction with the CS shell.\\
{\em Right.} Model velocity of the CDS ({\em thick line}) and the boundary velocity of the undisturbed supernova ejecta compared to the maximal velocity in the blue wing of absorption 
in \ha\ and  H$\beta$ ({\em empty diamonds}) and He\,I 10830\,\AA\ ({\em filled diamonds}). 
{\em Inset} shows CS density distribution.
 }
\label{fig:blc1}
\end{figure}

\clearpage

\begin{figure}
	\centering
	\includegraphics[width=0.97\columnwidth]{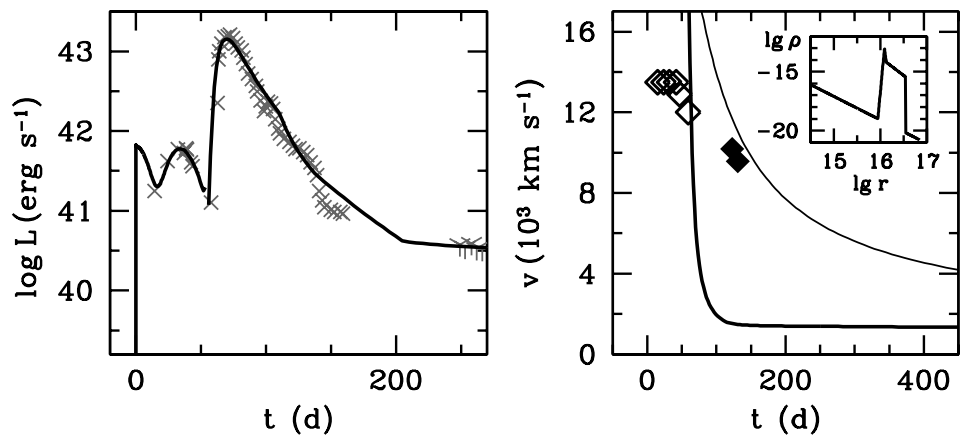}
	\caption{\rm
      The same as in Figure 1, but for the apherical model (nsm, Table 1).
  	}
	\label{fig:blc1}
\end{figure}

\end{document}